\documentclass[aps,showpacs,twocolumn,superscriptaddress,prl,floatfix]{revtex4}

\usepackage{graphicx}
\usepackage{dcolumn}
\usepackage{epstopdf}
\usepackage{setspace}

\begin{document}
\newcommand{\ttbar}{\mbox{$t\bar{t}$}}
\newcommand{\thetastar}{\mbox{$\theta^*$}}
\newcommand{\costhetastar}{\mbox{$\cos\thetastar$}}
\newcommand{\Mlb}{\mbox{$M_{lb}$}}
\newcommand{\Et}{\mbox{$E_T$}}
\newcommand{\Pt}{\mbox{$p_T$}}
\newcommand{\Flong}{\mbox{$F_0$}}
\newcommand{\Fplus}{\mbox{$F_+$}}
\newcommand{\met}{\mbox{$\protect \raisebox{0.3ex}{$\not$}\Et$}}

\bibliographystyle{revtex}

\title{Measurement of the Helicity of $W$ Bosons in Top-Quark Decays}
\affiliation{Institute of Physics, Academia Sinica, Taipei, Taiwan 11529, Republic of China} 
\affiliation{Argonne National Laboratory, Argonne, Illinois 60439} 
\affiliation{Institut de Fisica d'Altes Energies, Universitat Autonoma de Barcelona, E-08193, Bellaterra (Barcelona), Spain} 
\affiliation{Baylor University, Waco, Texas  76798} 
\affiliation{Istituto Nazionale di Fisica Nucleare, University of Bologna, I-40127 Bologna, Italy} 
\affiliation{Brandeis University, Waltham, Massachusetts 02254} 
\affiliation{University of California, Davis, Davis, California  95616} 
\affiliation{University of California, Los Angeles, Los Angeles, California  90024} 
\affiliation{University of California, San Diego, La Jolla, California  92093} 
\affiliation{University of California, Santa Barbara, Santa Barbara, California 93106} 
\affiliation{Instituto de Fisica de Cantabria, CSIC-University of Cantabria, 39005 Santander, Spain} 
\affiliation{Carnegie Mellon University, Pittsburgh, PA  15213} 
\affiliation{Enrico Fermi Institute, University of Chicago, Chicago, Illinois 60637} 
\affiliation{Joint Institute for Nuclear Research, RU-141980 Dubna, Russia} 
\affiliation{Duke University, Durham, North Carolina  27708} 
\affiliation{Fermi National Accelerator Laboratory, Batavia, Illinois 60510} 
\affiliation{University of Florida, Gainesville, Florida  32611} 
\affiliation{Laboratori Nazionali di Frascati, Istituto Nazionale di Fisica Nucleare, I-00044 Frascati, Italy} 
\affiliation{University of Geneva, CH-1211 Geneva 4, Switzerland} 
\affiliation{Glasgow University, Glasgow G12 8QQ, United Kingdom} 
\affiliation{Harvard University, Cambridge, Massachusetts 02138} 
\affiliation{Division of High Energy Physics, Department of Physics, University of Helsinki and Helsinki Institute of Physics, FIN-00014, Helsinki, Finland} 
\affiliation{University of Illinois, Urbana, Illinois 61801} 
\affiliation{The Johns Hopkins University, Baltimore, Maryland 21218} 
\affiliation{Institut f\"{u}r Experimentelle Kernphysik, Universit\"{a}t Karlsruhe, 76128 Karlsruhe, Germany} 
\affiliation{High Energy Accelerator Research Organization (KEK), Tsukuba, Ibaraki 305, Japan} 
\affiliation{Center for High Energy Physics: Kyungpook National University, Taegu 702-701; Seoul National University, Seoul 151-742; and SungKyunKwan University, Suwon 440-746; Korea} 
\affiliation{Ernest Orlando Lawrence Berkeley National Laboratory, Berkeley, California 94720} 
\affiliation{University of Liverpool, Liverpool L69 7ZE, United Kingdom} 
\affiliation{University College London, London WC1E 6BT, United Kingdom} 
\affiliation{Massachusetts Institute of Technology, Cambridge, Massachusetts  02139} 
\affiliation{Institute of Particle Physics: McGill University, Montr\'{e}al, Canada H3A~2T8; and University of Toronto, Toronto, Canada M5S~1A7} 
\affiliation{University of Michigan, Ann Arbor, Michigan 48109} 
\affiliation{Michigan State University, East Lansing, Michigan  48824} 
\affiliation{Institution for Theoretical and Experimental Physics, ITEP, Moscow 117259, Russia} 
\affiliation{University of New Mexico, Albuquerque, New Mexico 87131} 
\affiliation{Northwestern University, Evanston, Illinois  60208} 
\affiliation{The Ohio State University, Columbus, Ohio  43210} 
\affiliation{Okayama University, Okayama 700-8530, Japan} 
\affiliation{Osaka City University, Osaka 588, Japan} 
\affiliation{University of Oxford, Oxford OX1 3RH, United Kingdom} 
\affiliation{University of Padova, Istituto Nazionale di Fisica Nucleare, Sezione di Padova-Trento, I-35131 Padova, Italy} 
\affiliation{University of Pennsylvania, Philadelphia, Pennsylvania 19104} 
\affiliation{Istituto Nazionale di Fisica Nucleare Pisa, Universities of Pisa, Siena and Scuola Normale Superiore, I-56127 Pisa, Italy} 
\affiliation{University of Pittsburgh, Pittsburgh, Pennsylvania 15260} 
\affiliation{Purdue University, West Lafayette, Indiana 47907} 
\affiliation{University of Rochester, Rochester, New York 14627} 
\affiliation{The Rockefeller University, New York, New York 10021} 
\affiliation{Istituto Nazionale di Fisica Nucleare, Sezione di Roma 1, University of Rome ``La Sapienza," I-00185 Roma, Italy} 
\affiliation{Rutgers University, Piscataway, New Jersey 08855} 
\affiliation{Texas A\&M University, College Station, Texas 77843} 
\affiliation{Istituto Nazionale di Fisica Nucleare, University of Trieste/\ Udine, Italy} 
\affiliation{University of Tsukuba, Tsukuba, Ibaraki 305, Japan} 
\affiliation{Tufts University, Medford, Massachusetts 02155} 
\affiliation{Waseda University, Tokyo 169, Japan} 
\affiliation{Wayne State University, Detroit, Michigan  48201} 
\affiliation{University of Wisconsin, Madison, Wisconsin 53706} 
\affiliation{Yale University, New Haven, Connecticut 06520} 
\author{A.~Abulencia}
\affiliation{University of Illinois, Urbana, Illinois 61801}
\author{D.~Acosta}
\affiliation{University of Florida, Gainesville, Florida  32611}
\author{J.~Adelman}
\affiliation{Enrico Fermi Institute, University of Chicago, Chicago, Illinois 60637}
\author{T.~Affolder}
\affiliation{University of California, Santa Barbara, Santa Barbara, California 93106}
\author{T.~Akimoto}
\affiliation{University of Tsukuba, Tsukuba, Ibaraki 305, Japan}
\author{M.G.~Albrow}
\affiliation{Fermi National Accelerator Laboratory, Batavia, Illinois 60510}
\author{D.~Ambrose}
\affiliation{Fermi National Accelerator Laboratory, Batavia, Illinois 60510}
\author{S.~Amerio}
\affiliation{University of Padova, Istituto Nazionale di Fisica Nucleare, Sezione di Padova-Trento, I-35131 Padova, Italy}
\author{D.~Amidei}
\affiliation{University of Michigan, Ann Arbor, Michigan 48109}
\author{A.~Anastassov}
\affiliation{Rutgers University, Piscataway, New Jersey 08855}
% ...
\collaboration{CDF Collaboration}
\noaffiliation

\date{\today}

\begin{abstract}
We measure the branching fraction of the top quark to longitudinally and right-handed polarized $W$ bosons, $\Flong$ and $\Fplus$, using approximately 200~pb$^{-1}$ of $\bar{p}p$ collisions collected by the CDF II detector. We analyze two quantities sensitive to the $W$ helicity: the invariant mass of the charged lepton and the bottom-quark jet in the decay $t\rightarrow Wb \rightarrow \ell\nu b$ (where $\ell=e$ or $\mu$), and the transverse momentum of the charged lepton. We find $\Flong=0.74^{+0.22}_{-0.34}$, and $\Fplus<0.27$ at the 95\% confidence level. These measurements are in agreement with the standard model predictions. 
\end{abstract}

\maketitle

The top quark
 is the most massive known elementary fermion, with $m_t \sim 175$~GeV/$c^2$ \cite{topmass-CDF, topmass-D0}. At the Fermilab Tevatron proton-antiproton collider, with a center-of-mass energy of $\sqrt{s}=1.96$~TeV, most top quarks are pair-produced via the strong interaction \cite{topprod-theory1, topprod-theory2}. However, the decay $t\rightarrow Wb$ proceeds entirely via the weak interaction. 
Given the $V-A$ structure of the weak interaction in the standard model (SM), in the limit of a massless bottom quark the top quark can decay to either a left-handed or longitudinally-polarized $W^+$ boson \cite{charge-conj} and a bottom quark. The fraction $\Flong$ of longitudinally-polarized $W$ bosons is enhanced due to the large coupling of the top quark to the Higgs field responsible for electroweak symmetry-breaking. The leading-order SM prediction is \cite{whel-theory}

\begin{equation}
\label{eqn:F0}
   \Flong \equiv \frac{\Gamma(t\rightarrow W_0b)}{\Gamma(t\rightarrow W_0b)+\Gamma(t\rightarrow W_{\pm} b)}  = \frac{m_t^2}{2M_W^2 + m_t^2},
\end{equation}
where $W_0$ and $W_{\pm}$ indicate longitudinally- and transversely-polarized $W$'s respectively, and $M_W=80.4$~GeV/$c^2$ is the $W$ boson mass \cite{Wmass-ref}. For $m_t=175$~GeV/$c^2$, $\Flong=0.70$. A deviation from this prediction could indicate non-SM physics such as large CP-violation in top-quark decays \cite{non-SM}, as could a nonzero value for the right-handed fraction $\Fplus$.

We use two observables in \ttbar~candidate events to measure the $W$ helicity. The first is the decay angle \thetastar\ of the charged lepton in the $W$ decay frame, measured with respect to the top-quark direction, and the second is the transverse momentum \Pt\  of the charged lepton. Leptons from longitudinally-polarized $W$ boson decays have a symmetric angular distribution $\propto(1-\cos^2\thetastar)$, while left-handed $W$ decays have an asymmetric distribution $\propto(1-\costhetastar)^2$. We can approximate \costhetastar\ by relating it to the invariant mass of the system composed of the $b$ quark and the charged lepton, \Mlb:

\begin{equation}
\label{eqn:cstar-def}
\costhetastar = \frac{p_\ell\cdot p_b - E_\ell E_b}{|\mathbf{p}_\ell||\mathbf{p}_b|} \simeq\frac{2\Mlb^2}{m_t^2-M_W^2}-1,
\end{equation}
a variable that depends only on lab-frame momenta.
\begin{figure}
\begin{center}
\includegraphics[height=1.5in]{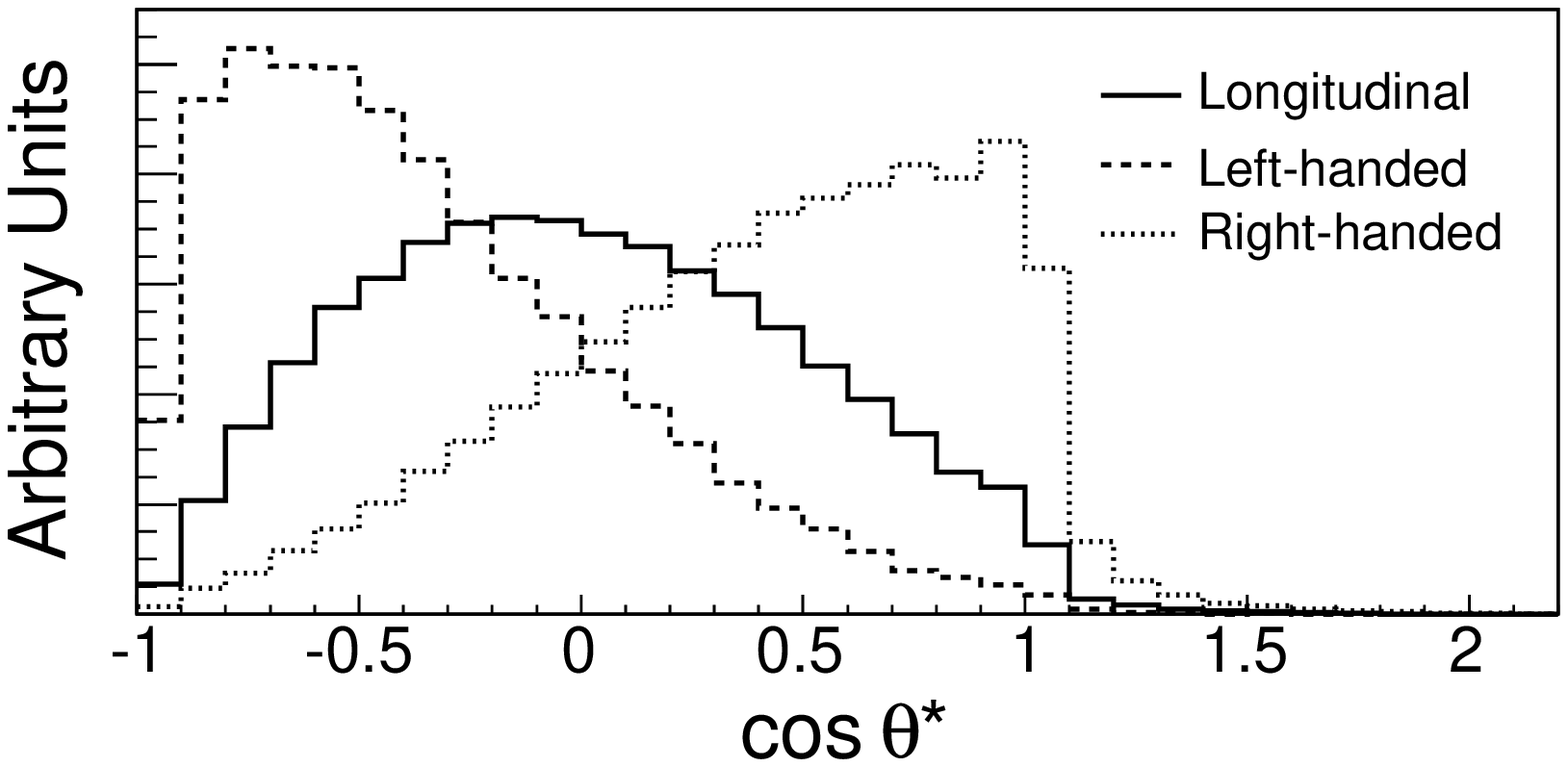}
\includegraphics[height=1.5in]{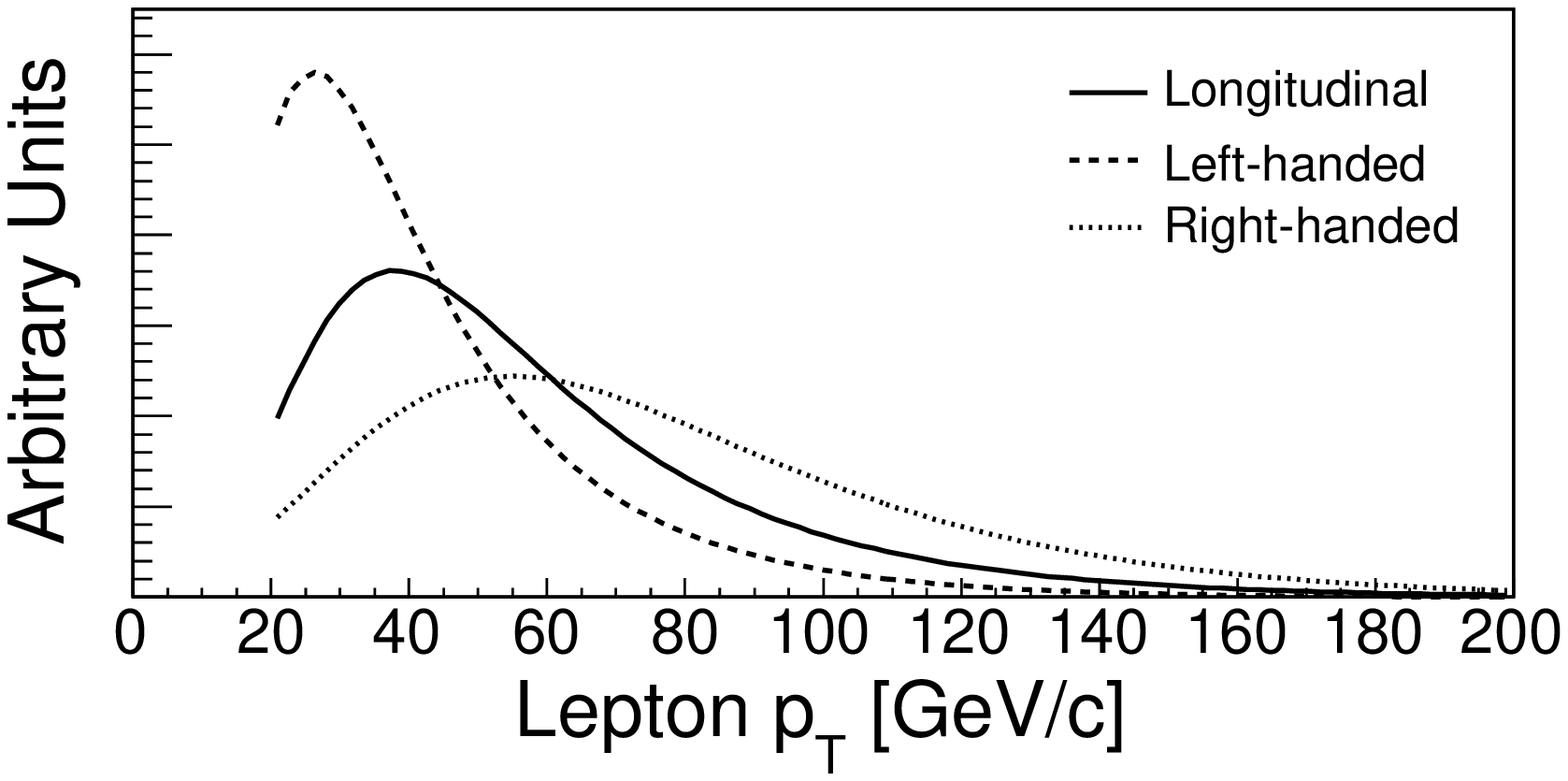}
\caption{\label{fig:templates}Distributions of reconstructed \costhetastar\ (upper plot) and lepton \Pt\ (lower) for top-quark decays to left-handed, right-handed,  and longitudinally polarized $W$ bosons.}
\end{center}
\end{figure}
The second observable, the charged lepton \Pt, exploits the fact that charged leptons from left-handed $W$ decays are preferentially emitted in the backward direction with respect to the $W$ direction of motion, leading to a softer \Pt\ in the lab frame, while the leptons from right-handed $W$'s are preferentially emitted forward and thus have a harder \Pt\ spectrum. Longitudinal $W$ decays represent an intermediate case. Figure~\ref{fig:templates} shows the predicted \costhetastar\ and lepton \Pt\ distributions for $m_t=175$~GeV/$c^2$, after the event selection and reconstruction described below. 

A measurement of $\Flong$ has been previously reported by the CDF Collaboration \cite{CDF-Winn} using $\approx100$~pb$^{-1}$ of data from the 1992-1996 Tevatron collider run (Run I). Using the $\Pt$ technique, a value of $0.91\pm 0.37\mathrm{(stat)}\pm 0.13\mathrm{(syst)}$ was obtained. Using the same data set, CDF has also placed a limit on the right-handed helicity fraction of $\Fplus<0.18$ at the 95\% confidence level (C.L.) with the \Mlb\ technique \cite{CDF-Kilminster}. The D\O~Collaboration has used 125~pb$^{-1}$ of Run I data to obtain $\Flong = 0.56\pm 0.31$ \cite{D0-Florencia}.
Here we report a measurement of $\Flong$ and $\Fplus$  that combines the \Mlb\ and $\Pt$ techniques. 

The CDF II detector \cite{CDFII-det} consists of a charged-particle tracking system in a magnetic field of 1.4~T, segmented electromagnetic and hadronic calorimeters, and muon detectors. A silicon microstrip detector provides tracking over the radial range 1.5-28~cm and is used to detect displaced secondary vertices. The fiducial region of the silicon detector covers the pseudo-rapidity range $|\eta| < 2$, while the central tracking system and muon chambers provide coverage for $|\eta| < 1$ \cite{CDF-coord}.  For electron identification we use the calorimeter region $|\eta|<1$, while for jet identification we use $|\eta|<2.5$. A three-level trigger system selects events with electron (muon) candidates with $\Et$ $(\Pt ) > 18$ GeV (18 GeV/$c$), which form the data set for this analysis. 

 In the decay process $\ttbar\rightarrow W^+bW^-\bar{b}$, events can be classified based on the observed number of isolated charged leptons with large transverse momentum, where a lepton signifies an electron or muon of either charge; typically these leptons come from the decay $W\rightarrow\ell\nu$. Transverse momentum for electrons from $W$ decay is best measured at CDF using the transverse energy $E_T$ deposited in the calorimeter, while for muons the transverse momentum \Pt\ is measured by the tracking system.  We will use the symbol $\Pt$ to denote the appropriate calorimeter- or tracking-based quantity. The 193~pb$^{-1}$ ``dilepton" sample \cite{dil-PRD} consists of events with two oppositely-charged lepton candidates, each  with $\Pt > 20$ GeV/$c$. Events in this sample are required to have missing transverse energy $\met > 25$~GeV, and two or more jets with pseudorapidity $|\eta|<2.5$ and transverse energy $E_T>15$~GeV.  The scalar sum of the transverse energy of the jets, leptons, and $\met$, is required to be greater than 200~GeV.  We observe 13 events in this sample, with a predicted total background from $WW$ pairs, $Z\rightarrow\bar{\tau}\tau$, the Drell-Yan process, and hadrons misidentified as leptons (``fakes") of $2.7\pm 0.7$ events. The 162~pb$^{-1}$ ``lepton plus jets" sample \cite{ljet-xsec} consists of events with a single isolated lepton candidate with $\Pt > 20$~GeV$/c$, $\met > 20$~GeV, and three or more jets with $|\eta|<2$ and $E_T>15$ GeV. To reduce backgrounds, we require that one or more jets have a displaced secondary-vertex tag, indicating that it is consistent with the decay of a long-lived $b$ hadron. Fifty-seven events pass the selection cuts, of which approximately 2/3 are \ttbar\ events. The largest remaining backgrounds come from $W$ plus jets events containing bottom or charm jets, QCD multijet events, and $W$ plus light-quark events misidentified as $b$'s. 
 
The \Pt\ analysis \cite{Nate-thesis} uses both samples, while the \Mlb\ analysis \cite{Trevor-thesis} uses the lepton plus jets sample only.  In addition to the selection requirements described above, events selected for the \Mlb\ analysis are required to have a fourth jet with $\Et>8$~GeV and $|\eta|<2$. Thirty-seven events pass this cut. The presence of four jets allows the event to be kinematically reconstructed as a \ttbar\ event \cite{topmass-CDF} with the top mass constrained to 175 GeV/$c^2$, and to associate the appropriate jet to the lepton in Equation~\ref{eqn:cstar-def}.  We find that 31 of the 37 events pass a cut on the fit quality, with an estimated background of $6.9\pm 0.9$ events. 

To create reconstructed \costhetastar\ templates for \ttbar\ signal events, we use the \textsc{madevent} \cite{madevent}  Monte Carlo program. Hadronization and fragmentation are carried out using \textsc{pythia} \cite{pythia}. Events for the $\Pt$ analysis are generated using \textsc{herwig} \cite{herwig}. In both cases, we fix the helicity in the top rest frame  of one $W$ boson,  while the other $W$ takes on values according to the SM prediction.  The events are then passed through the CDF simulation and reconstruction algorithms. 
For the lepton plus jets sample, all backgrounds except QCD are modeled with Monte Carlo simulations. We model the QCD background using lepton plus jets events where the primary lepton is non-isolated.  For the dilepton sample all but the fake background is modeled with Monte Carlo. We model the latter background using lepton plus jet events containing jets that could be misidentified as a charged lepton.

\begin{figure}
 \begin{center}
 \includegraphics[height=3.0in]{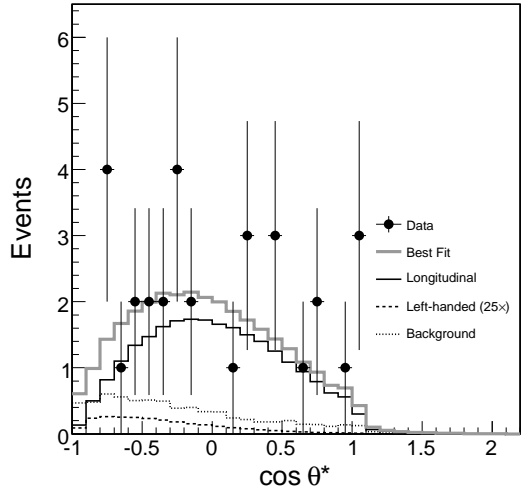}
 \caption{\label{fig:cstar-fit} The reconstructed \costhetastar\ distribution for the lepton plus jets sample, overlaid with signal and background templates according to their best-fit values. The left-handed template has been scaled up by a factor of 25. }
 \end{center}
 \end{figure}

\begin{figure}
 \begin{center}
 \includegraphics[height=3.0in]{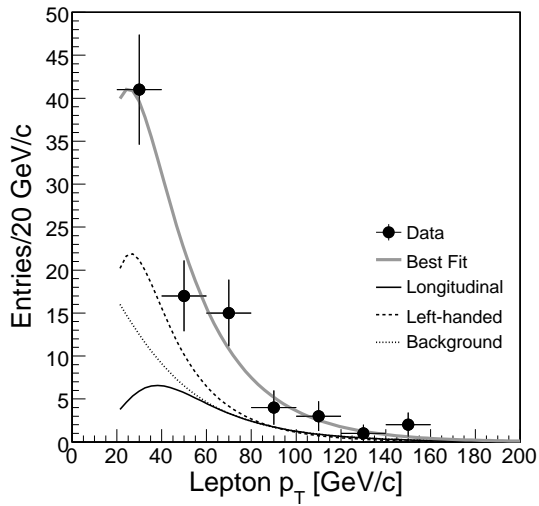}
 \caption{\label{fig:pt-fit}Distribution of lepton \Pt\ for the lepton plus jets and dilepton samples, overlaid with the total signal and background templates according to their best-fit values.}
 \end{center}
 \end{figure}

 The data are fit separately to the \costhetastar\ and \Pt\ templates using likelihood functions that include a Gaussian constraint on the background, as well as corrections for trigger and event selection cuts that have helicity-dependent biases, such as those on the lepton \Pt. Because the statistical power of the sample is insufficient to fit $\Fplus$ and $\Flong$ simultaneously, we constrain \Fplus\ to zero when fitting for \Flong; when fitting for $\Fplus$ we constrain $\Flong$ to 0.70. The remaining fit parameter is unconstrained. The results of the fits to the various subsamples are shown in Table~\ref{tab:results}. The reconstructed \costhetastar\ distribution from the data and the best-fit templates are shown in Figure~\ref{fig:cstar-fit}. The observed \costhetastar\ distribution extends somewhat beyond the physical range $-1\le\costhetastar\le 1$ because the world-average top and $W$ masses are used in Equation~\ref{eqn:cstar-def}, rather than the true event-by-event reconstructed masses, whose much larger uncertainties would unnecessarily smear the 
\costhetastar\ distribution obtained from the $M_{lb}$ approximation. In the dilepton sample, the best-fit value of $\Flong$ falls at $-0.54^{+0.35}_{-0.25}$, outside the physical range. In this case, the observed distribution of lepton \Pt\ is softer than any component of signal or background in our model. 
A measured central value of $-0.54$ or less is expected 0.5\% of the time for a true $\Flong$ of 0.7; moreover a previous analysis of the kinematics of the dilepton data \cite{ivanov} has found them 
 to be consistent with the SM at the 1.0-4.5\%  level.
We therefore carry out our \textit{a priori} decision to perform a combined \Pt\ fit to the two samples. The lepton \Pt\ distribution for the two samples and the results of the fit are shown in Figure~\ref{fig:pt-fit}.

 \begin{table}
 \begin{center}
   \caption{\label{tab:results} Summary of results for the \Mlb, \Pt, and combined measurements of \Flong\ and \Fplus. $N$ is the number of events or leptons used in the measurement. Where two uncertainties are given the first is statistical and the second is systematic. Uncertainties on the combined measurements are the total statistical and systematic uncertainty.}
   \begin{tabular}{cccccc}
   \hline\hline
   Analysis &$N$& \Flong & \Fplus  \\ \hline
   \Mlb & 31& $0.99^{+0.29}_{-0.35}\pm0.19$  & $0.23\pm0.16\pm0.08$  \\  
   \Pt(dilepton) & 26& $-0.54^{+0.35}_{-0.25}\pm 0.16$ & $-0.47\pm0.10\pm0.09$ \\
   \Pt(lep+jets) & 57& $0.95^{+0.35}_{-0.42}\pm0.17$  & $0.11^{+0.21}_{-0.19}\pm0.10$ \\
   \Pt(combined) & 83& $0.31^{+0.37}_{-0.23}\pm0.17$  & $-0.18^{+0.14}_{-0.12}\pm0.12$ \\ \hline
   Combined &$\cdots$ & $0.74^{+0.22}_{-0.34}$ &  $0.00^{+0.20}_{-0.19}$ \\
   95\% C.L. limit & $\cdots$ & {$<0.95$},{$>0.18$} & $<0.27$ \\
   
   \hline\hline
   \end{tabular}
   \end{center}
 \end{table}

 The dominant systematic uncertainties in the \Mlb\ and \Pt\ analyses arise from uncertainties in the top-quark mass, the background shape and normalization, the effects of initial- and final-state radiation (ISR/FSR), and the parton distribution functions (PDFs).
 We determine these uncertainties by performing Monte Carlo experiments in which the systematic parameter in question is varied by $\pm 1\sigma$ and the resulting simulated data are fit to the default templates.  We compare the mean $\Flong$ or \Fplus\ returned by the likelihood fit with the default (unfluctuated) value. The results are summarized in Table~\ref{tab:systematics}. The sum in quadrature of all sources of systematic uncertainty leads to a final result of $\Flong = 0.99^{+0.29}_{-0.35} (\rm stat.)\pm 0.19{\rm (syst.)}$ for the \Mlb\ analysis and  $\Flong = 0.31^{+0.37}_{-0.23}{\rm (stat.)}\pm0.17{\rm (syst.)}$ for the \Pt\ analysis. 
 
\begin{table}[h!]
 \begin{center}
 \caption{\label{tab:systematics}
    Summary of systematic uncertainties for the measurements of \Flong\ and \Fplus.}
  \begin{tabular}{c|c|c|c|c}
   \hline\hline
   	Systematic Source & \multicolumn{2}{c}{\Pt\ Method} \vline & \multicolumn{2}{c}{\Mlb\ Method}\\ 
	\cline{2-5}
	                        & $\Delta F_0$ & $\Delta F_+$ & $\Delta F_0$ & $\Delta F_+$\\ 
	\hline
	Top Mass 	          & 0.11  & 0.09     & 0.08      & 0.04\\
	Bkg. Modeling & 0.10  & 0.06     & 0.13      & 0.05\\
	ISR/FSR		  & 0.04  & 0.03     & 0.03      & 0.02\\
	PDF 		          & 0.03  & 0.03     & 0.04      & 0.01\\
	MC Statistics		          & 0.01  & $<$0.01  & 0.01      & 0.01\\
	Acceptance Correction	          & 0.02  & 0.01     & $<$ 0.005 & $<$ 0.005\\
	Trigger Correction                & 0.02  & 0.02     & $\cdots$        & $\cdots$\\
	Jet Energy Scale   & $\cdots$   & $\cdots$       & 0.09      & 0.04\\
	MC Modeling 		  & $\cdots$    & $\cdots$       & 0.04      & 0.02\\
	$b$-tagging 	          & $\cdots$    & $\cdots$       & 0.01      & $<$ 0.005\\
	\hline
	Total			          & 0.17  & 0.12     & 0.19      & 0.08\\
   \hline\hline
  \end{tabular}
 \end{center}	
\end{table}
 
We combine the results of the \Mlb\ and \Pt\ analyses taking into account both the statistical and systematic correlations between the two techniques. Statistical correlations arise because the two analyses share the subset of the lepton plus jets sample that passes the fit quality cut on the top mass reconstruction. Common sources of systematic uncertainty include the top mass uncertainty and background normalizations. The
correlation coefficients are determined via Monte Carlo experiments. The combined result is  $\Flong = 0.74^{+0.22}_{-0.34}$ (stat.+syst.). In addition, we find $\Fplus=0.00^{+0.20}_{-0.19}$ (stat.+syst.) and $\Fplus<0.27$ at the 95\% C.L.  These results are consistent with the SM predictions of $\Flong=0.70$, $\Fplus=0$.

 We thank Tim Stelzer and Fabio Maltoni for their help with the \textsc{madevent} calculations. We thank the Fermilab staff and the technical staffs of the participating institutions for their vital contributions. This work was supported by the U.S. Department of Energy and National Science Foundation; the Italian Istituto Nazionale di Fisica Nucleare; the Ministry of Education, Culture, Sports, Science and Technology of Japan; the Natural Sciences and Engineering Research Council of Canada; the National Science Council of the Republic of China; the Swiss National Science Foundation; the A.P. Sloan Foundation; the Bundesministerium f\"ur Bildung und Forschung, Germany; the Korean Science and Engineering Foundation and the Korean Research Foundation; the Particle Physics and Astronomy Research Council and the Royal Society, UK; the Russian Foundation for Basic Research; the Comisi\'on Interministerial de Ciencia y Tecnolog\'{\i}a, Spain; in part by the European Community's Human Potential Programme under contract HPRN-CT-2002-00292; and the Academy of Finland.


\begin{thebibliography}{99}
\bibitem{charge-conj} Charge-conjugation symmetry implies that the $\bar{t}$ quark decays to either
a longitudinally- or right-handed-polarized $W^-$. Throughout the remainder of this Letter, a ``left-handed $W$" refers to either a left-handed $W^+$ or a right-handed $W^-$.
\bibitem{CDF-coord} In the CDF geometry, $\theta$ is the polar angle with respect to the proton beam axis, and $\phi$ is the azimuthal angle. The pseudo-rapidity is $\eta\equiv -\ln(\tan(\theta/2))$. The transverse momentum $\Pt$ is the component of the momentum projected onto the plane perpendicular to the beam axis. The transverse energy $\Et$ of a shower or calorimeter tower is $E\sin\theta$, where $E$ is the energy deposited. $\met\equiv -\sum_i\Et_i\mathbf{n}_i$, where $\mathbf{n}_i$ is the unit vector in the azimuthal plane that points from the beam line to the $i$th calorimeter tower.
\end{thebibliography}
\end{document}